\documentclass[12pt]{article}
\usepackage{amsmath}
\usepackage{amssymb}
\usepackage{epsfig}
\usepackage{lscape}

\begin{document}

\begin{center}

{\Large \bf Parton distributions from deep-inelastic-scattering data}

\vspace{1cm}
{\bf S.~I.~Alekhin}

\vspace{0.1in}
{\baselineskip=14pt Institute for High Energy Physics, 142281 Protvino, Russia}

\begin{abstract}
We perform the analysis of existing light-targets
deep-inelastic-scattering (DIS) data
in the leading-order (LO), next-to-leading-order 
(NLO), and next-to-next-to-leading-order (NNLO) QCD approximations
and extract PDFs simultaneously with the value of the strong coupling constant 
$\alpha_s$ and the high-twist contribution to the structure functions.
The main theoretical uncertainties and experimental uncertainties 
due to all sources of experimental errors in data
are estimated, the latter generally dominate for the obtained PDFs.
The uncertainty in Higgs boson production cross section due to errors in PDFs 
is $\sim 2$\% for the LHC and 
varies from 2\% to 10\% for the Fermilab collider under variation of 
the Higgs boson mass from $100~{\rm GeV}$ to $300~{\rm GeV}$.
For the $W$-boson production cross section the uncertainty is 
$\sim 2$\% for the both colliders. The value of 
$\alpha^{\rm NNLO}_{\rm s}(M_{\rm Z})=0.1143\pm 0.0014({\rm exp.})$
is obtained, while the high-twist terms 
do not vanish up to the NNLO as required by comparison to data.
\end{abstract}
\end{center}
{\bf PACS numbers:} 13.60.Hb,06.20.Jr,12.38.Bx\\
{\bf Keywords:} deep inelastic scattering, parton distributions, 
strong coupling constant, high twists

\newpage

\section {Introduction} 

The deep-inelastic-scattering (DIS) data are used for extraction of  
the parton distribution functions (PDFs) in nucleon starting 
from the advent of the Bjorken scaling phenomena (for the most 
recent studies see 
Refs.~\cite{Barone:1999yv,Botje:1999dj,Alekhin:2001ch}). 
However since new DIS data 
appear permanently and the theoretical basis for their analysis 
is being developed alongside, those analysis should be updated in order to 
improve the experimental and theoretical errors on the extracted PDFs.  
In this paper we give such update of the 
next-to-leading-order (NLO) QCD analysis of 
Ref.~\cite{Alekhin:2001ch} aimed to extract PDFs, 
the high-twist (HT) contribution to the 
structure functions $F_{2,\rm L}$, and the value of strong coupling
constant $\alpha_{\rm s}$
from the existing data on DIS of charged leptons off
the proton and deuterium targets. 
One of the improvements of our analysis is replacement
of data by the H1 \cite{Aid:1996au} and 
the ZEUS \cite{Derrick:1996hn} collaborations 
by their more recent data \cite{Adloff:2000qk,Chekanov:2001qu} 
obtained in the 1996-97 run of the HERA collider, when  
these experiments accumulated about 5 times more events than before.
Another improvement is account of the NNLO QCD corrections, which
have been known with a good precision after calculation of   
the Mellin moments of the NNLO splitting functions up to 
14-th~\cite{Retey:2000nq}. This new input significantly decreases the 
uncertainty of PDFs due to the higher-orders (HO) QCD 
corrections~\cite{vanNeerven:2000wp,Alekhin:2001ih} and 
together with more precise data allows to 
produce the NLO and NNLO PDFs sets with reduced total uncertainty.

The particular features of the analysis \cite{Alekhin:2001ch}
as compared to the global fits of Refs.\cite{Pumplin:2002vw,Martin:2001es}
are:
\begin{itemize}

\item data on the differential cross sections are used that allows to 
use experimental constraints on the structure function $F_{\rm L}$;

\item corrections for the effects of target mass (TM) and for the nuclear 
effects in deuteron are applied to the data, the both were calculated 
in the fits iteratively using the current PDFs;    

\item comprehensive account of the experimental errors and 
their correlations is performed;

\item impact of main sources of the theoretical errors is analyzed.

\end{itemize}

All these features are also kept in the present analysis.

\begin{landscape}
\begin{table}
\caption{The numbers of data points (NDP) for 
separate experiments and the corresponding renormalization shifts $\xi$ 
obtained in different fits.}
\begin{center}
\begin{tabular}{ccccccccc} 
&\multicolumn{4}{c}{proton} &\multicolumn{4}{c}{deuterium}  
\\ \cline{2-5} \cline{6-9}
Experiment&NDP&\multicolumn{3}{c}{$\xi$[\%]}&NDP&\multicolumn{3}{c}{$\xi$[\%]} 
\\ \cline{3-5} \cline{7-9}
&&LO&NLO&NNLO&&LO&NLO&NNLO \\ \hline
SLAC-E-49A  &58 &$2.9 \pm 1.4$&$1.6\pm1.4$&$1.1\pm1.4$ &58 &$0.0 \pm 1.3$&$-1.3\pm1.3$&$-1.7\pm1.3$   \\  
SLAC-E-49B  &144&$3.2 \pm 1.4$&$1.8\pm1.4$&$1.2\pm1.4$&135&$0.5 \pm 1.4$&$-0.8\pm1.4$&$-1.3\pm1.4$  \\
SLAC-E-87   &90&$2.9 \pm 1.4$&$1.8\pm1.4$&$1.2\pm1.4$&90&$0.7 \pm 1.3$&$-0.5\pm1.3$&$-0.9\pm1.3$   \\
SLAC-E-89A   &66&$5.6 \pm 1.9$&$4.0\pm1.9$&$3.1\pm1.9$&59&$2.3 \pm 2.0$&$0.5\pm2.0$&$-0.3\pm2.0$  \\
SLAC-E-89B  &79&$2.6 \pm 1.3$&$1.2\pm1.4$&$0.7\pm1.4$&62&$-0.2 \pm 1.3$&$-1.4\pm1.3$&$-1.9\pm1.4$   \\
SLAC-E-139   &&&&&16&$1.6 \pm 1.3$&$0.2\pm1.3$&$-0.2\pm1.3$  \\
SLAC-E-140  &&&&&26&&&   \\
BCDMS  &351&&&&254&&& \\ 
NMC(90 GeV)    &46&$-1.5 \pm 1.6$&$-1.4\pm1.5$&$-1.3\pm1.6$&46&$-2.6 \pm 1.5$&$-3.0\pm1.5$&$-2.9\pm1.5$  \\
NMC(120 GeV)    &59&$-0.2 \pm 1.5$&$0.2\pm1.5$&$0.4\pm1.5$&59&$-1.3 \pm 1.5$&$-1.6\pm1.4$&$-1.5\pm1.5$ \\
NMC(200 GeV)    &66&$2.1 \pm 1.5$&$1.9\pm1.4$&$2.0\pm1.4$&66&$0.0 \pm 1.4$&$-0.3\pm1.4$&$-0.3\pm1.4$ \\
NMC(280 GeV)    &74&$1.3 \pm 1.4$&$0.9\pm1.4$&$0.8\pm1.4$&74&$-1.1 \pm 1.4$&$-1.4\pm1.4$&$-1.4\pm1.4$ \\
H1    &135&&&&&&& \\
ZEUS  &161&&&&&&& \\ 
\end{tabular}
\end{center}
\label{tab:ndp}
\normalsize
\end{table}
\end{landscape}

\section{Theoretical and experimental input of the fit}

The model for data description is based on perturbative QCD 
with phenomenological parameterization of the LT and
HT contributions to the structure functions $F_{\rm 2,L}$.
The analysis was performed in the ${\overline{\rm MS}}$ 
scheme with the number of flavors 
fixed at 3.
The LT PDFs are parameterized at $Q^2_0=9~{\rm GeV}^2$ in the following form:
\begin{equation}
xu_{\rm V}(x,Q_0)=\frac{2}{N^{\rm V}_{\rm u}}
x^{a_{\rm u}}(1-x)^{b_{\rm u}}(1+\gamma_2^{\rm u}x),
\label{eqn:pdf1}
\end{equation}
\begin{equation}
xu_{\rm S}(x,Q_0)=\frac{A_{\rm S}}{N_{\rm S}}
\eta_{\rm u} x^{a_{\rm s}}(1-x)^{b_{\rm su}},
\end{equation}
\begin{equation}
xd_{\rm V}(x,Q_0)=\frac{1}{N^{\rm V}_{\rm d}}x^{a_{\rm d}}(1-x)^{b_{\rm d}},
\end{equation}
\begin{equation}
xd_{\rm S}(x,Q_0)=\frac{A_{\rm S}}{N^{\rm S}}x^{a_{\rm s}}(1-x)^{b_{\rm sd}},
\end{equation}
\begin{equation}
xs_{\rm S}(x,Q_0)=\frac{A_{\rm S}}{N^{\rm S}}\eta_{\rm s}
x^{a_{\rm s}}(1-x)^{(b_{\rm su}+b_{\rm sd})/2},
\end{equation}
\begin{equation}
xG(x,Q_0)=A_{\rm G}x^{a_{\rm G}}(1-x)^{b_{\rm G}}
(1+\gamma^{\rm G}_1\sqrt{x}+\gamma^{\rm G}_2 x),
\label{eqn:pdf2}
\end{equation}
where $u,d,s,G$ are the up, down, strange quarks,
and gluons distributions respectively; 
indices $V$ and $S$ correspond to the valence 
and sea quarks. The parameters $N^{\rm V}_{\rm u}, N^{\rm V}_{\rm d}$  and 
$A_{\rm G}$ were calculated 
from the other parameters using conservation of the partons momentum 
and the fermion number. The 
normalization parameter $N^{\rm S}$ is selected in such way that 
$A_{\rm S}$ gives the total momentum carried by the sea quarks.  
The parameter $\eta_{\rm s}$ was fixed at 0.42, in agreement with 
the results of NuTeV collaboration~\cite{Adams:1999sx}. 
The rest of parameters 
coming to Eqns.(\ref{eqn:pdf1}--\ref{eqn:pdf2}) were fitted to the data,
we convinced that no extra parameters are needed to improve the 
data description.
The HT contributions to the fitted structure functions $F_{\rm 2,L}$ were
parameterized in additive form
$$
F_{\rm 2,L}=F_{\rm 2,L}^{\rm LT,TMC}+\frac{H_{\rm 2,L}(x)}{Q^2},
$$
where $F_{\rm 2,L}^{\rm LT,TMC}$ are the LT terms with account of the 
TM correction\footnote{See 
Ref.~\protect\cite{Alekhin:2001ch} for detailed formula.}
and functions $H(x)$ are parameterized in the piece-linear form
and fitted to the data.

\begin{table}
\caption{The values of $\chi^2/{\rm NDP}$ for different fits 
and of the averages of residuals $R$ for the NLO fit
together with their standard deviations $\Delta R$
for separate data sets used in the analysis.}
\begin{center}
\begin{tabular}{cccccc} 
&NDP& \multicolumn{3}{c}{$\chi^2$/NDP} &$R(\Delta R)$ \\ \cline{3-5} 
Experiment& &LO&NLO&NNLO &\\  \hline
SLAC-E-49A  &115   &0.54&0.52&0.52   &$-0.04(0.23)$\\  
SLAC-E-49B  &279 &1.27&1.20&1.21    &$0.23(0.29)$ \\  
SLAC-E-87   &180   &0.96&0.92&0.91   &$0.01(0.37)$\\  
SLAC-E-89A   &125  &1.35&1.33&1.36   &$-0.18(0.45)$ \\  
SLAC-E-89B  &141   &0.94&0.82&0.81   &$0.47(0.49)$\\  
SLAC-E-139   &16  &0.81&0.58&0.62   &$-0.07(0.43)$ \\  
SLAC-E-140  &26   &1.31&0.94&0.90   &$0.17(0.86)$ \\  
BCDMS  &605 &1.15&1.13&1.12        &$-0.06(0.68)$\\  
NMC    &490 &1.35&1.24&1.24    &$-0.06(0.39)$\\   
H1(96-97)    &135 &1.39&0.97&1.09   &$0.42(0.55)$\\   
ZEUS(96-97)    &161 &1.54&1.32&1.28   &$-0.55(0.64)$\\   \hline
TOTAL  &2274 &1.20&1.11&1.11     &$0.01(0.22)$\\ 
\end{tabular}
\label{tab:chi}
\end{center}
\end{table}

The code used to evolve these distributions was used earlier in 
the analysis of Ref.~\cite{Alekhin:2001ih}.
This code was checked against Les Houches benchmark of Ref.\cite{Giele:2002hx}
and demonstrated accuracy much better than experimental precision 
of the analyzed data. 
A brief description of the data sets used in the analysis 
is given in Table~\ref{tab:ndp}.
We cut data with $Q^2<2.5~{\rm GeV}^2$, $Q^2>300~{\rm GeV}^2$,
and $x>0.75$ in order to 
minimize effects of the higher-order QCD corrections, 
electroweak interference, and uncertainties in 
the TM and deuteron corrections correspondingly.
All experimental uncertainties in data released by 
authors were accounted for 
in the analysis, including statistical errors and
correlated systematical errors
(in particular the general normalization errors).

\begin{figure}
\centerline{\epsfig{file=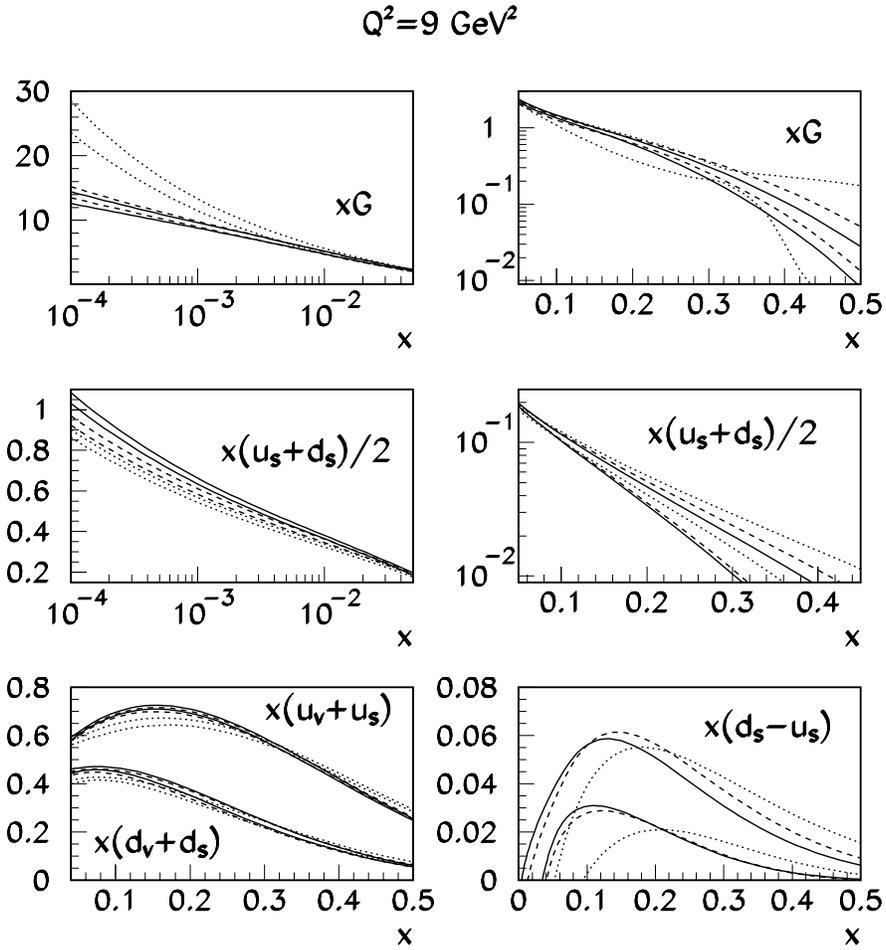,width=14cm,height=14cm}}
\caption{The experimental (statistical and systematical) errors bands for 
the PDFs obtain in the LO (dots), NLO (dashes), and NNLO (full)
fits.}
\label{fig:pdf}
\end{figure}

The total normalization of the NMC and most of the SLAC data subsets used in 
our analysis were not determined in the experiments, instead they were
fitted by authors
to the results of other experiments. Since now we have more DIS data 
than it was used in those fits, we perform 
independent renormalization of the NMC and SLAC data subsets that allows 
for more precise determination of the corresponding normalization factors.
Following the approaches adopted by those
experimental groups we introduce an additional 
normalization parameter for each beam energy and target in the NMC data set 
and for each experiment and target in the SLAC data sets.
The data from SLAC-E-140 experiment, which was performed 
with particular attention paid to the 
accurate determination of the absolute normalization, were not renormalized
-- we accounted for its general normalization
error given by the authors in the same way as the other 
correlated systematic errors.
The same approach was used for treatment of the general normalization 
error in the BCDMS, H1, and ZEUS data.
The normalization factors for the SLAC and NMC data
are given in Table~\ref{tab:ndp}
for the fits performed in the LO, NLO, and NNLO approximations.
In the LO fit the shifts for the SLAC deuterium sets
deviate off zero by about two standard deviations,
while for the NNLO fit they vanish, i.e.
account of the higher-order corrections 
decreases tension between the different data sets
(evidently, this effect is less pronounced for the NMC data than for the SLAC 
one since the later correspond to 
the lower values of $Q$, where these corrections are more important).
In average, renormalization of the SLAC data sets in the NLO fit is 
$\sim +2$\% for the proton and $\sim -0.5$\% for the deuterium targets.
For the NMC data the corresponding values also are different and, besides,
vary with the beam energy that justifies application
of the flexible renormalization scheme for this experiment.
Note that in particular due to this flexible renormalization we obtain 
satisfactory description of the NMC data in the NLO:
$\chi^2/{\rm NDP}$=1.24 in contrast with value 1.5 
obtained in the fit of Ref.~\cite{Pumplin:2002vw}. 
The values of $\chi^2/{\rm NDP}$ obtained in different orders of QCD are given 
in Table~\ref{tab:chi}. One can see that quality of the fit is approximately
the same for the NLO and NNLO fits and much worse for the LO fit.

\begin{figure}
\centerline{\epsfig{file=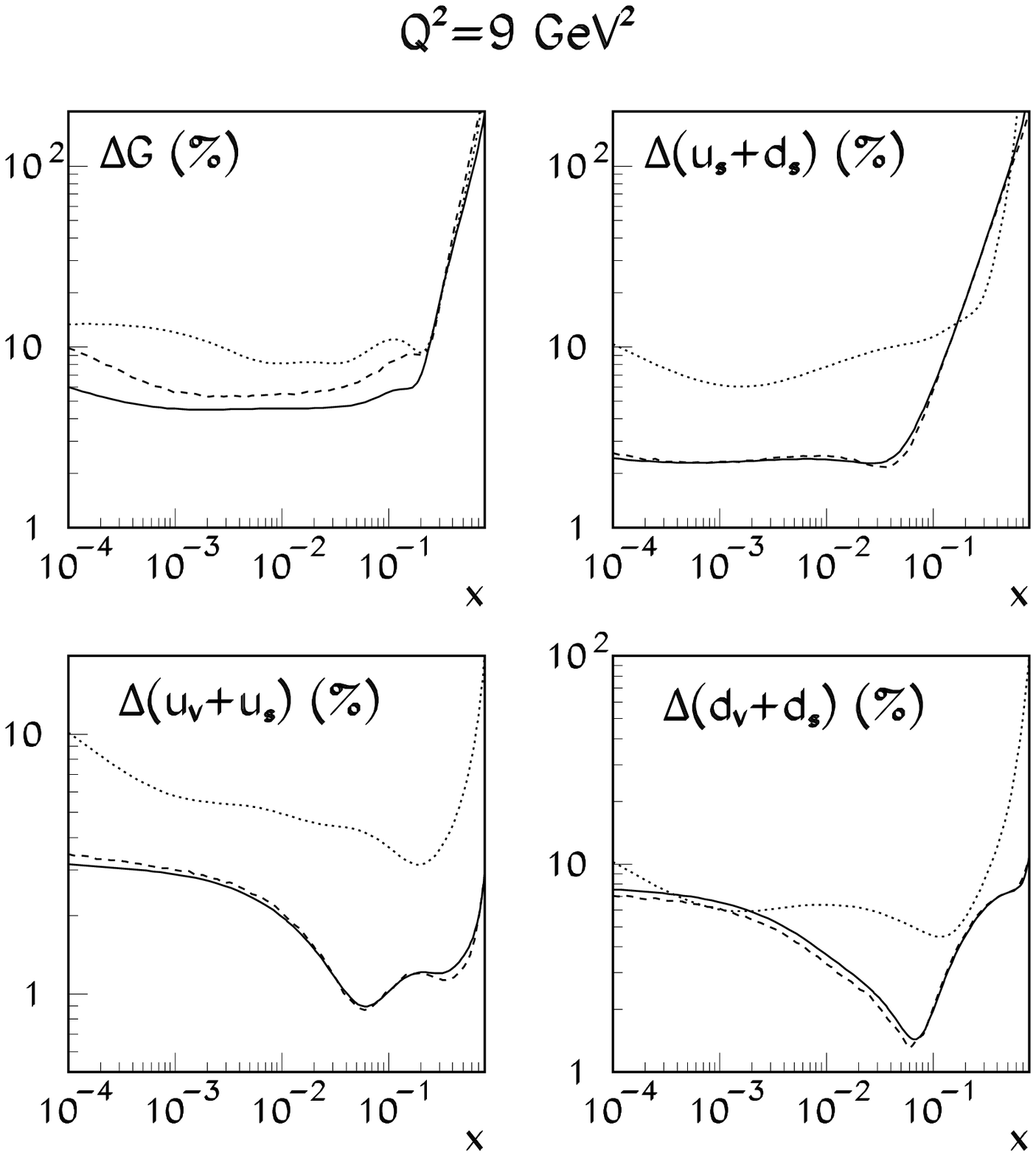,width=14cm,height=12cm}}
\caption{Impact of changes in the data set on the NLO PDFs and their errors. 
Full lines: relative experimental errors 
in our PDFs; dashes: the same for the PDFs of  
Ref.~\protect{\cite{Alekhin:2001ch}}; dots: the same for the CTEQ6 PDFs.}
\label{fig:err}
\end{figure}

\section {PDFs and their experimental errors}
\label{sec:pdfexp}

\begin{figure}
\centerline{\epsfig{file=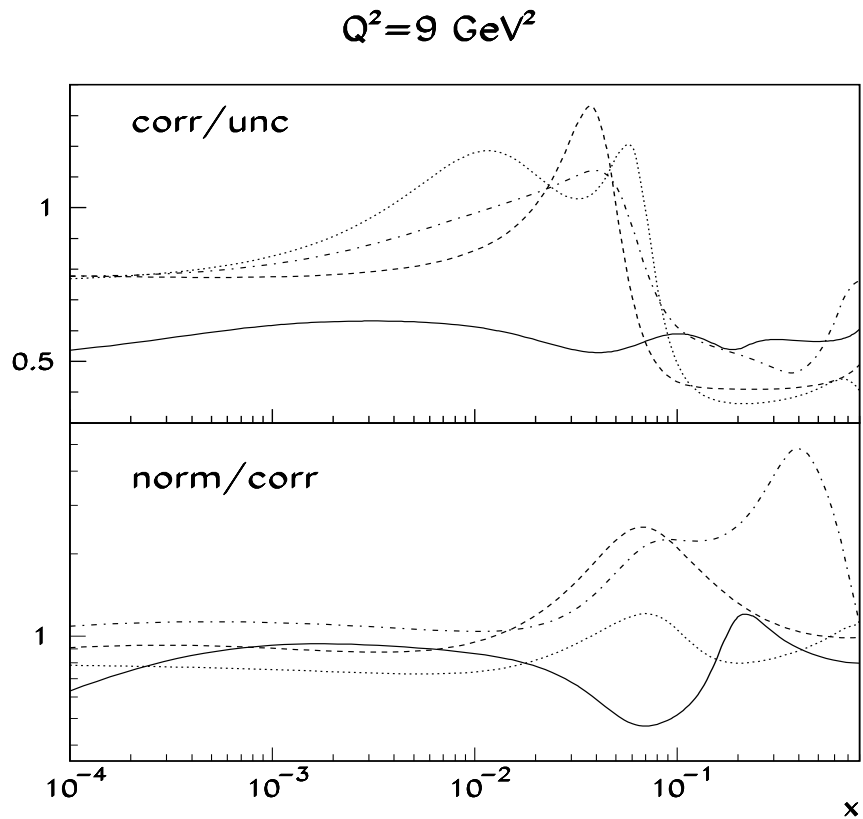,width=14cm,height=12cm}}
\caption{Ratios of the correlated systematic errors to the uncorrelated ones
(upper panel) and of the normalization errors to the correlated ones
(lower panel) for different PDFs (full curves: gluons;
dashes: sea quarks; dots: $d$-quarks, dashed-dots: $u$-quarks).}
\label{fig:pdfsys}
\end{figure}

The PDFs obtained from the fits in different pQCD orders 
with their total experimental uncertainties are 
given in Fig.\ref{fig:pdf}, the PDFs parameters 
are given in Table~\ref{tab:pars}.
One can see that the shift between 
the NLO and NNLO PDFs is generally within their experimental errors
excluding the sea quarks distribution at small $x$. The LO 
gluon distribution at small $x$ and valence quarks distributions
at moderate $x$ are very different from the corresponding NLO and NNLO PDFs.
This difference cannot be completely ascribed to the re-definition of 
PDFs in different orders since data are poorly described by pQCD in the LO.
Fitted LO PDFs are evidently  
distorted due to inaccount of the HO corrections and this distortion 
should be taken into account in calculation of other 
hard processes in the LO approximation. The PDFs errors 
consecutively fall from LO to NNLO due to additional terms 
of the perturbative expansions put additional constraints on 
the fitted values. This effect is most pronounced for the gluon distribution 
at large $x$. The comparison with experimental uncertainties 
in the NLO PDFs obtained from our earlier analysis of
Ref.~\cite{Alekhin:2001ch} is given in Fig.\ref{fig:err}. The main 
improvement is in precision of the gluon distribution 
at small $x$ due to more precise HERA data have been used.
The total experimental errors can be conventionally split into 3 
components: the uncorrelated, general normalization, 
and correlated errors (excluding normalization ones).
The normalization errors can also be considered as correlated ones, but 
we separate them for the purpose of comparison 
since very often the normalization errors are accounted 
for in a peculiar way or not accounted for at all.
In our analysis the normalization errors for BCDMS, H1, and ZEUS data sets 
are included in the general covariance matrix, while for SLAC and NMC
data sets they appear through the error matrix of the fitted parameters. 
Ratios of the correlated errors to the uncorrelated ones
and of the normalization errors to the correlated ones for different PDFs
are given in Fig.\ref{fig:pdfsys}. One can see that for most of PDFs
the correlated errors are more important at small $x$. The normalization 
errors are largest at large $x$; in this region 
they give main contribution to the total experimental uncertainties in PDFs. 

\begin{landscape}
\begin{table}
\caption{The values of the fitted parameters of the PDFs.} 
\begin{center}
\begin{tabular}{cccccc}   
  &     &LO  & NLO                      &\multicolumn{2}{c}{NNLO}\\ \cline{5-6}
  &     &     &                          &(A+B)/2& B \\\hline
Valence &&&&& \\
       &$a_{\rm u}$&$0.551\pm0.085$ &$0.700\pm0.030$&$0.725\pm0.027$  &$0.725\pm0.026$\\
       &$b_{\rm u}$&$3.672\pm0.042  $ &$3.920\pm0.050$&$4.024\pm0.050$  &$4.011\pm0.050$\\
       &$\gamma_2^{\rm u}$&$3.0\pm1.7  $ &$1.14\pm0.39$&$1.05\pm0.35$  &$1.01\pm0.34$\\
       &$a_{\rm d}$&$0.639\pm0.047  $ &$0.722\pm0.082$&$0.772\pm0.074$      &$0.818\pm0.062$\\
       &$b_{\rm d}$&$4.48\pm0.23  $ &$4.94\pm0.12$&$5.14\pm0.15$         &$5.22\pm0.18$\\
Glue   &&&& &\\
       &$a_{\rm G}$&$-0.302\pm0.021  $ &$-0.146\pm0.018$&$-0.128\pm0.021$   &$-0.096\pm0.020$\\
       &$b_{\rm G}$&$5.3\pm5.1  $ &$8.2\pm1.3$&$9.4\pm1.1$     &$9.8\pm1.0$\\
       &$\gamma_1^{\rm G}$&$-1.94\pm0.83  $ &$-3.76\pm0.45$&$-3.84\pm0.52$   &$-4.18\pm0.45$\\
       &$\gamma_2^{\rm G}$&$2.8\pm5.3  $ &$7.7\pm1.7$&$8.6\pm1.7$   &$9.2\pm1.5$\\
Sea    &&&& &\\
       &$A_{\rm S}$&$0.161\pm0.013  $ &$0.160\pm0.011$&$0.1591\pm0.0090$&$0.1545\pm0.0071$\\
       &$a_{sd}$& $-0.1980\pm0.0057  $&$-0.1968\pm0.0048$&$-0.2092\pm0.0044$&$-0.2086\pm0.0043$\\
       &$b_{\rm sd}$&$3.8\pm1.1  $ &$5.1\pm1.4$&$5.6\pm1.3$&$6.5\pm1.1$\\
       &$\eta_{\rm u}$&$1.46\pm0.15  $ &$1.16\pm0.12$&$1.12\pm0.11$&$1.036\pm0.099$\\
       &$b_{\rm su}$&$9.2\pm1.8  $ &$10.16\pm0.93$&$10.49\pm0.90$&$10.43\pm0.89$\\
       &&&& &\\
\end{tabular}
\end{center}
\label{tab:pars}
\end{table}
\end{landscape}

\begin{figure}
\centerline{\epsfig{file=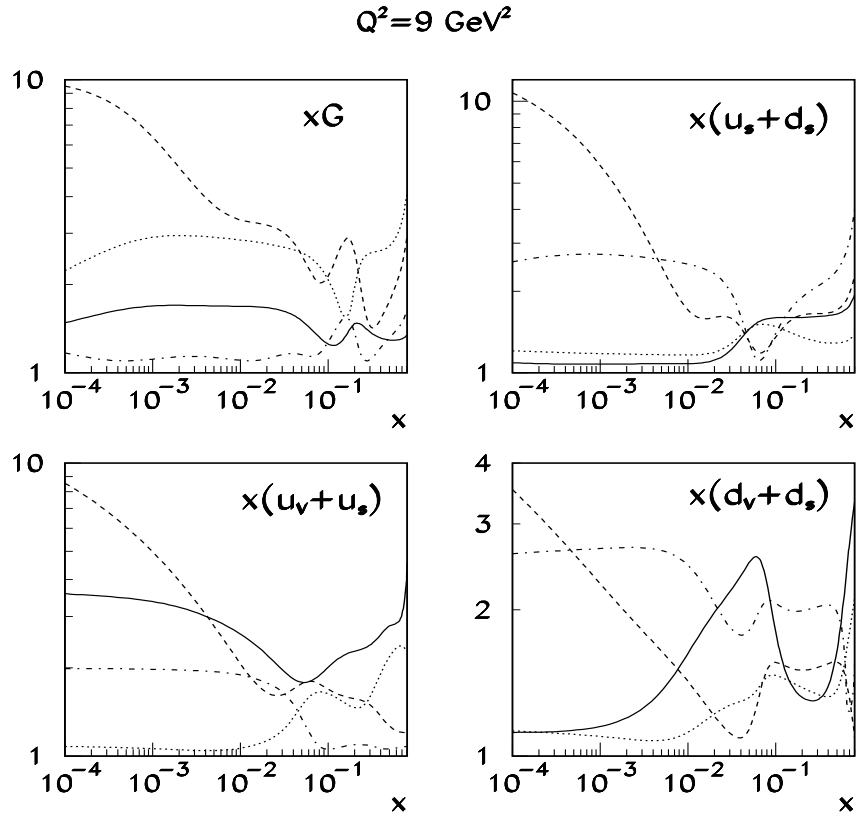,width=14cm,height=12cm}}
\caption{Relative increase of the experimental errors in our NLO PDFs 
due to rejection of different data sets from the fit 
(full curves: BCDMS data are rejected; dashes: HERA ones; dots: SLAC ones; 
dashed-dots: NMC ones).}
\label{fig:excl}
\end{figure}

Impact of each data set on the precision of our PDFs
is demonstrated in Fig.\ref{fig:excl}. The HERA data are crucial for 
determination of the gluon and quarks distributions at small $x$;
the BCDMS data constraint the quark distributions at large x;
the NMC data are essential for determination of the anti-quarks 
and $d$-quark distributions at large $x$; and the SLAC data 
improve precision of the $d$-quark and gluon distributions at large x. 
In summary none of the subsets
can be dropped out without deterioration of the PDFs precision.
We also checked how sensitive are the errors in PDFs to 
the inclusion in fit the ZEUS data of Ref.~\cite{Chekanov:2001qu}
with $Q^2>250~{\rm GeV}^2$, the H1 data of 
Ref.~\cite{Adloff:1999ah} with $Q^2>150~{\rm GeV}^2$,
and the FNAL-E-665 data of Ref.~\cite{Adams:1996gu}. If one applies to these 
data our regular kinematical cuts $Q^2>2.5~{\rm GeV}^2$ and $x<0.75$,
the relative suppression of the PDFs errors after each of 
these data sets has been included does not exceed 0.9 and therefore they  
are useless for the PDFs determination.

\begin{figure}
\centerline{\epsfig{file=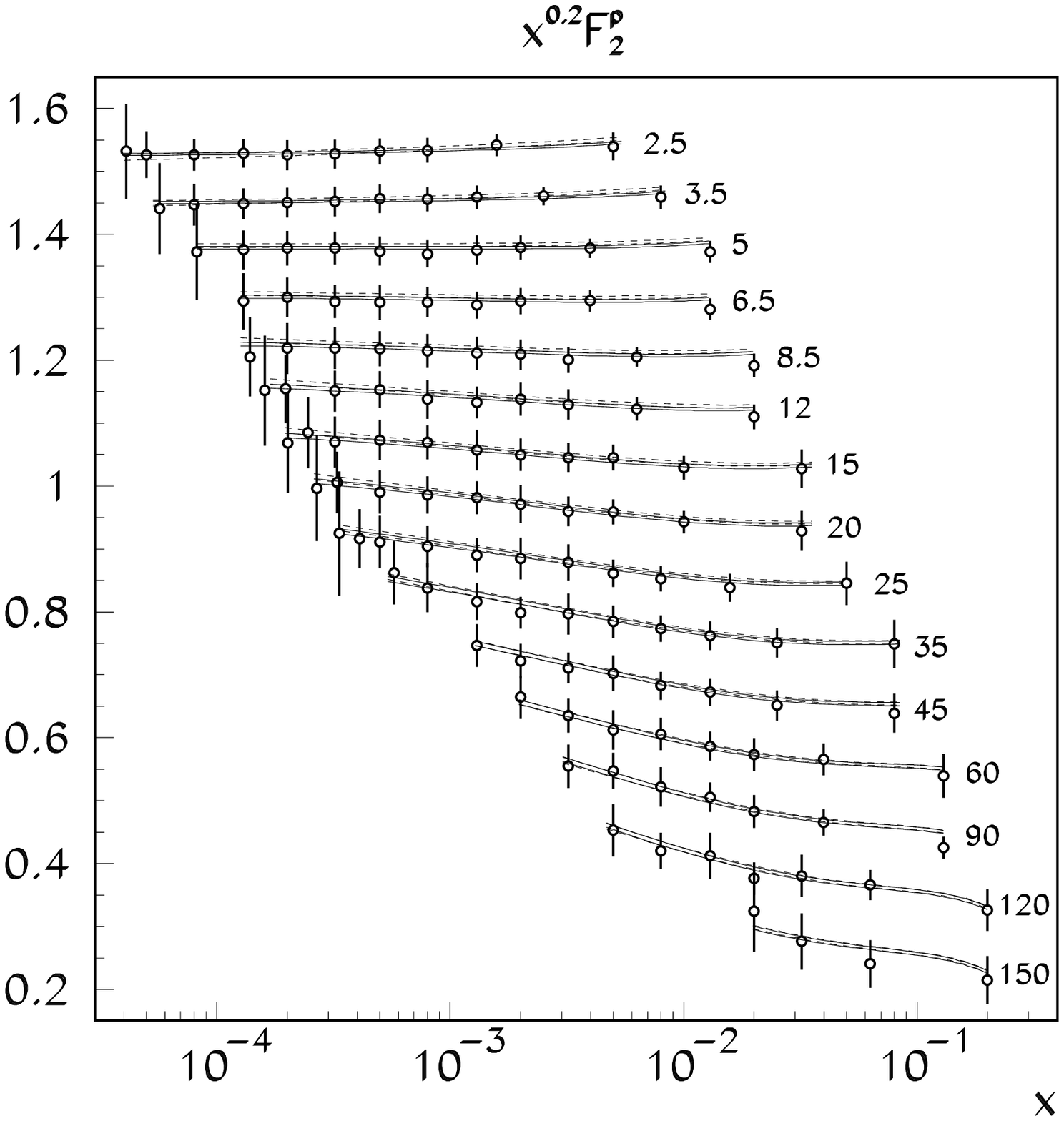,width=14cm,height=14cm}}
\caption{The experimental error bands for $F_2^p$ calculated 
using our NLO PDFs (full curves) compared to the 
H1 data used in the fit.
Figures at the curves are values of $Q^2$ in units of GeV$^2$. 
For better view the factor of  
$x^{0.2}$ and vertical shifts are applied to the data points and curves.
The same bands calculated using the A99 PDFs are also given 
for comparison (dashed curves).}
\label{fig:h1}
\end{figure}
\begin{figure}
\centerline{\epsfig{file=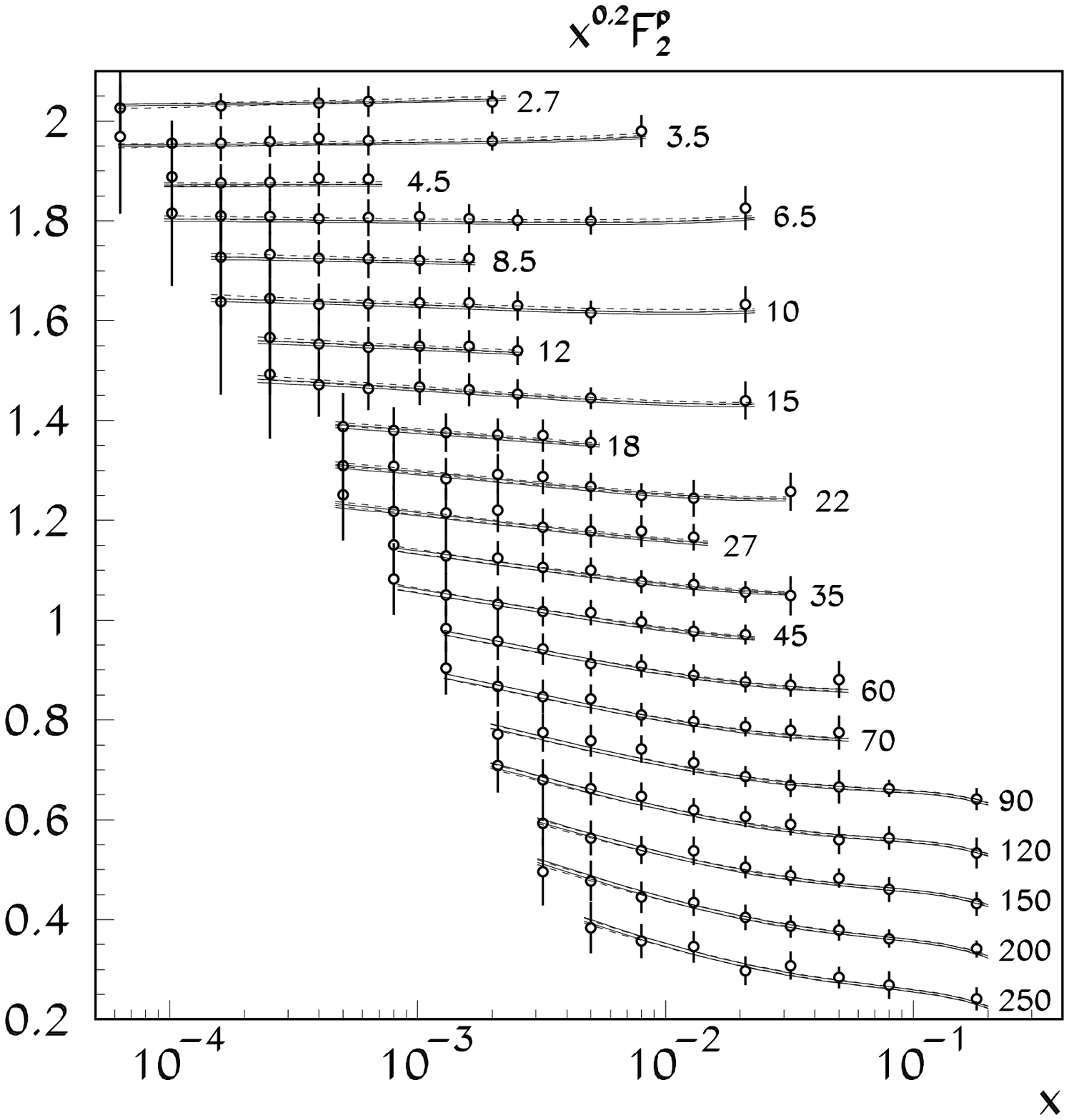,width=14cm,height=14cm}}
\caption{The same as Fig.\protect\ref{fig:h1} for the ZEUS data 
used in the fit.}
\label{fig:zeus}
\end{figure}

\begin{figure}
\centerline{\epsfig{file=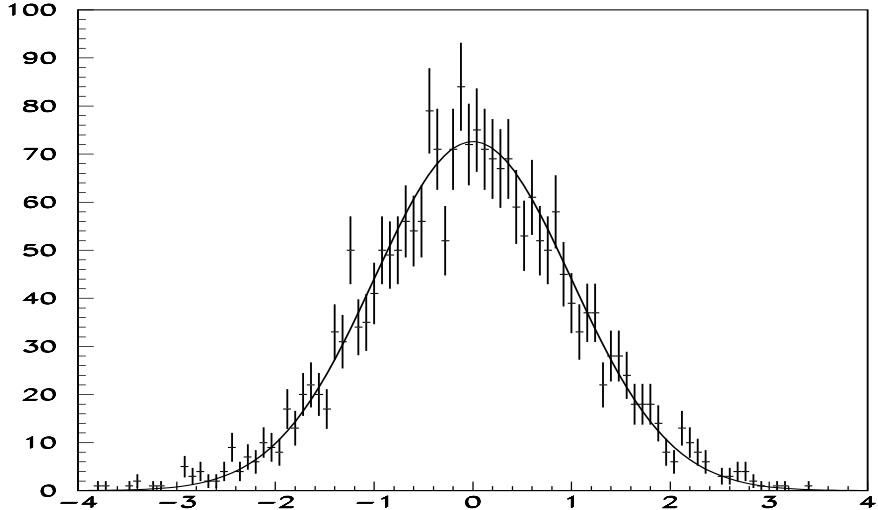,width=14cm,height=8cm}}
\caption{Distribution of the diagonalized residuals in our NLO fit.
The curve superimposed gives normal distribution.}
\label{fig:res}
\end{figure}

The novel HERA data used in our fit is confronted with the fitted curves
in Figs.\ref{fig:h1},\ref{fig:zeus}. 
For comparison we show in the same figures the calculations 
based on our earlier PDFs of Ref.\cite{Alekhin:2001ch}. The 
experimental error bands for both sets of PDFs overlap that proves stability 
of the analysis. If the fit is unbiased, 
the fitted curve and data should coincide modulo 
statistical fluctuations. These fluctuations are quantified by 
the average of residuals $R=<(f-y)/e>$, 
where $f$ is the fitted function, $y$ is 
measurement, and $e$ is the total experimental error in data. 
The standard deviation of this average $\Delta R$
is $\sim 1/\sqrt{NDP}$ for data set with uncorrelated points.
If data are correlated, coherent shift of the fitted curve
within the value of common systematic error in data is allowed
and value of $\Delta R$ gets larger, up to 1 for the case of strong 
correlations. Therefore the 
value of $\Delta R$ can be used as a crude indicator of 
significance of correlations in data set.
In our analysis $\Delta R$ is largest for the BCDMS and SLAC-E-140 data
(see Table~\ref{tab:chi}). Since 
magnitudes of $R$ are well within $\Delta R$ both for all subsets and 
for the total set we conclude that our fit is unbiased.
Despite the systematic errors are not necessarily Gaussian distributed 
the value of $\chi/NDP=1.1$ obtained in our fit proves that this 
is good approximation in our case. Distribution of the
diagonalized residuals (DRs), which are the 
components of the vector of residuals $(f-y)/e$
multiplied by the square root of the inverse covariance 
matrix is plotted in Fig.\ref{fig:res}. If the errors obey the Gaussian 
distribution the distribution of DRs have to be normal~\cite{Alekhin:2000es}
and as one can see in Fig.\ref{fig:res} this is the case for our fit. 
Therefore all errors are Gaussian in good approximation and 
consequently the errors in fitted parameters are Gaussian 
too\footnote{The width of the DRs distribution is 1.05 
that is marginally larger than 
the width of the normal one; for this reason rescaling 
of the errors in spirit of the PDG's procedure would lead to 
negligible effect.}. 

\section {The PDFs theoretical uncertainties}

With experimental errors in PDFs permanently reduced due to 
to more relevant data appear the theoretical errors become more 
dominating contribution to the total PDFs uncertainty.
In study of the importance of theoretical uncertainties
the experimental errors serve as a natural scale.
The variations below this scale can be neglected and, moreover, 
they just can occur due to statistical fluctuations in data.
For this reason we analyzed only the theoretical errors,
which give effect larger or comparable to the experimental ones.

\subsection{Uncertainty due to higher order QCD corrections}

\begin{figure}
\centerline{\epsfig{file=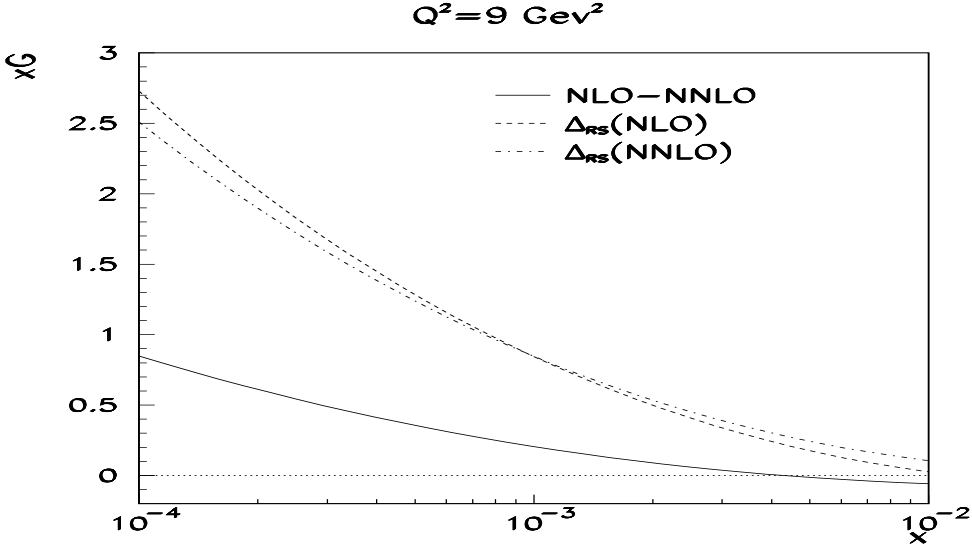,width=14cm,height=8cm}}
\caption{Perturbative stability of the gluon distribution at small $x$. 
The corresponding RS error $\Delta_{RS}$ in the NLO and NNLO is given for
comparison.}
\label{fig:pdf_order}
\end{figure}

\begin{figure}
\centerline{\epsfig{file=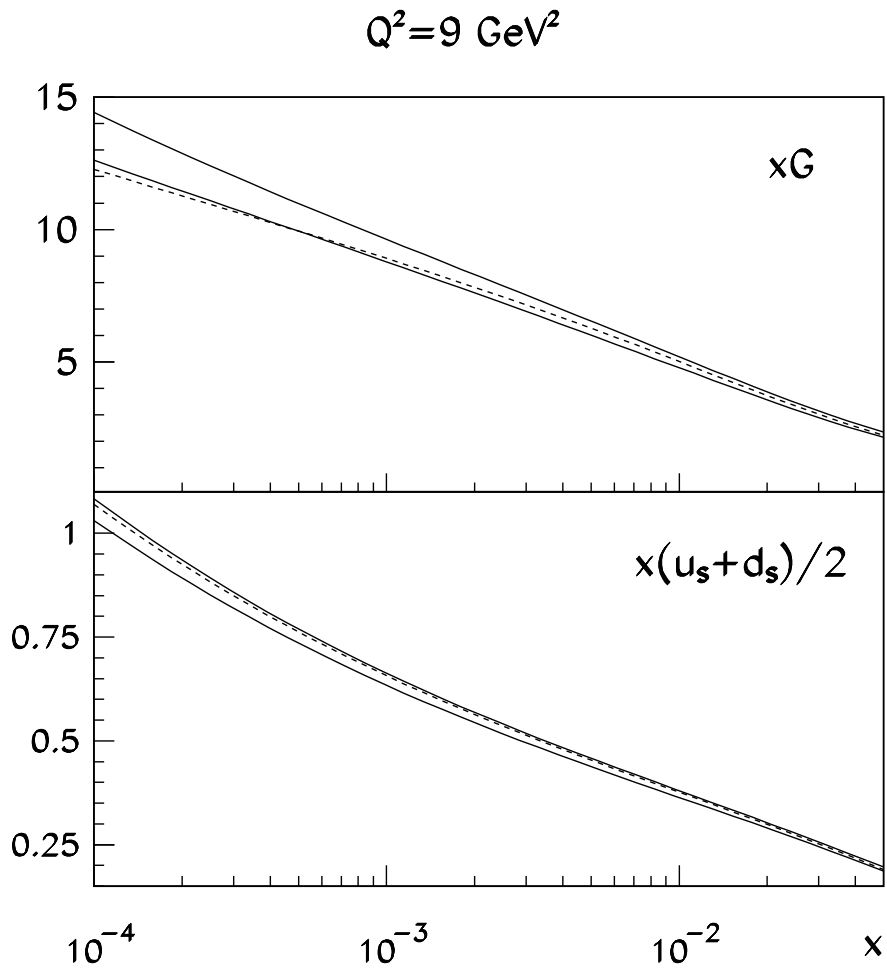,width=14cm,height=14cm}}
\caption{The NNLO PDFs obtained with the different choices of the 
NNLO kernel (full lines: 1$\sigma$ experimental bands 
for the nominal fit; dashes: fit with the NNLO kernel modified within 
the allowed uncertainty given in Ref.\protect\cite{vanNeerven:2000wp}).}  
\label{fig:pdfnnlo}
\end{figure}

The uncertainty in PDFs due to the HO QCD corrections
is not an uncertainty in full sense because the 
PDFs differ from order to order just by definition. 
Nevertheless the variation of 
PDFs with the pQCD order is of practical importance
for estimation of the HO QCD uncertainties in the hard cross sections,
which are calculated using the PDFs, despite the variation of the 
cross section value due to the HO corrections to PDFs
can be in some cases partially compensated by  
the HO corrections to the corresponding coefficient functions.
A common approach used for estimation of the HO errors 
in the analysis of DIS data is to vary the QCD renormalization scale (RS) 
and factorization scale (FS) from $Q^2$ to $4Q^2$.
Evidently, since the range of this variation is 
arbitrary and no $x$-dependence of the scale factors
is assumed, this approach gives only semi-qualitative estimate on the HO 
uncertainty.
Having the PDFs at different pQCD orders we can verify this approach
comparing the real variation of PDFs from order to order 
with the HO errors estimated using the variation of the pQCD scales.
Example of such comparison is given in Fig.\ref{fig:pdf_order}.
One can see that the real change of the gluon 
distribution from the NLO to the NNLO is much smaller than 
the RS error at small $x$. With the FS errors taken into account the 
difference would get even larger. Adjustment of the range of the scale 
variation would also not help since the signs of
the NLO-NNLO difference and the RS error are opposite at $x\sim 0.01$. 
Note that the RS error in gluon distribution almost does not 
decrease from the NLO to the NNLO. This means that 
if the perturbative expansion does converge the discrepancy between 
the NNLO RS error and the change of gluon distribution 
from the NNLO to the NNNLO is even larger. For this reason 
we recommend to use the difference between NLO and NNLO PDFs 
as estimate of the HO uncertainty motivated by the 
properties of the asymptotic expansions.
The NNLO PDFs have an additional source of errors
due to possible variation of the NNLO kernel. 
The magnitude of these errors is largest at small $x$, but even in this region 
they are smaller than the experimental errors (see Fig.\ref{fig:pdfnnlo}).

\subsection{Uncertainties due to the heavy quarks contribution}

\subsubsection{Variation of the factorization scheme}
\label{sec:VFN}

\begin{figure}
\centerline{\epsfig{file=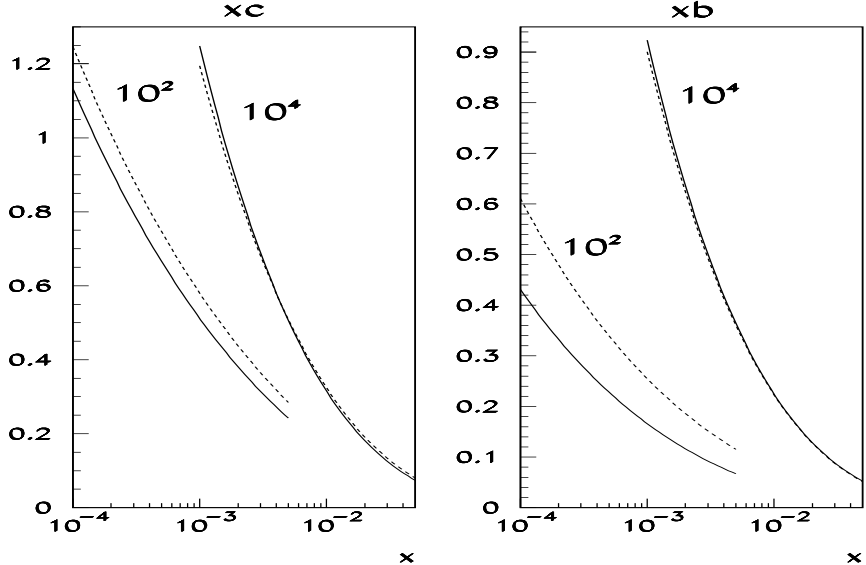,width=14cm,height=8cm}}
\caption{The $c$- and $b$-quarks distributions in the VFN scheme 
(dashed curves) compared to the corresponding FFN 
contributions to the structure function $F_2$
weighted with the quarks charges (full curves).
The figures at the curves give the values of $Q^2$ in GeV$^2$;
the range of $x$ for a given $Q^2$ is typical for the processes  
studied at the Fermilab and LHC colliders.}
\label{fig:vfn}
\end{figure}

In our analysis contribution of the $c$- and $b$-quarks was calculated 
in the fixed-flavor-number (FFN) scheme using 
the photon-gluon fusion mechanism~\cite{Witten:bh} with account of the  
NLO QCD correction~\cite{Laenen:1992xs}. Since 
the heavy-quark contribution rises with $Q$ rapidly 
the FFN scheme without resummation of large logs
is irrelevant
at small $x$ and $Q$ much larger than the quark mass~\cite{Shifman:1977yb}. 
In order to overcome this difficulty
the variable-flavor-number (VFN) scheme was proposed
\cite{Collins:mp}. In the VFN scheme
the heavy quarks are considered as massless and are included 
in the QCD evolution at $Q=m_{c,b,t}$, where $m_{c,b,t}$ are the masses
of $c$-, $b$-, and $t$-quark correspondingly.
The VFN scheme does include resummation of large logs, but 
at the same time evidently overestimates the heavy-quarks distributions
at $Q$ close to the thresholds. Nevertheless this scheme
is widely used in practice 
due to it allows for to greatly simplify the calculation of processes with 
the heavy quarks involved. The FFN scheme is appropriate  
in our case since the heavy-quarks contribution 
is concentrated at small $x$, while the small $x$ data used in our analysis 
correspond to the values of $Q$, which are not very far from the 
threshold of heavy quarks production. Nevertheless in order  
to lighten implementation of our PDFs in different calculations
we generated the VFN PDFs using our FFN PDFs as input. For this purpose
we evolved our PDFs from our boundary value of $Q_0^2=9~{\rm GeV}^2$ 
to $Q=m_c$ in the FFN scheme and used the obtained PDFs 
as the boundary condition for the VFN evolution.
The obtained VFN distributions of $c$- and $b$-quarks in the NLO are given 
in Fig.\ref{fig:vfn} in comparison with
the corresponding contributions to the structure function $F_2$
calculated in the FFN scheme and weighted for the quarks charges. 
At large $Q$ the FFN scheme gives results similar to the VFN ones
at the values of 
$x$ feasible by future hadron colliders. At low $Q$ the VFN 
scheme gives sizeable overestimation of the heavy-quark distributions, 
especially for the $b$-quarks. For this reason the certain caution is 
necessary in use of the VFN PDFs 
at small $Q$ (see also Ref.~\cite{Gluck:1993dp} in this connection). 

\subsubsection{Uncertainty due the error in heavy quarks masses}
    
\begin{figure}
\centerline{\epsfig{file=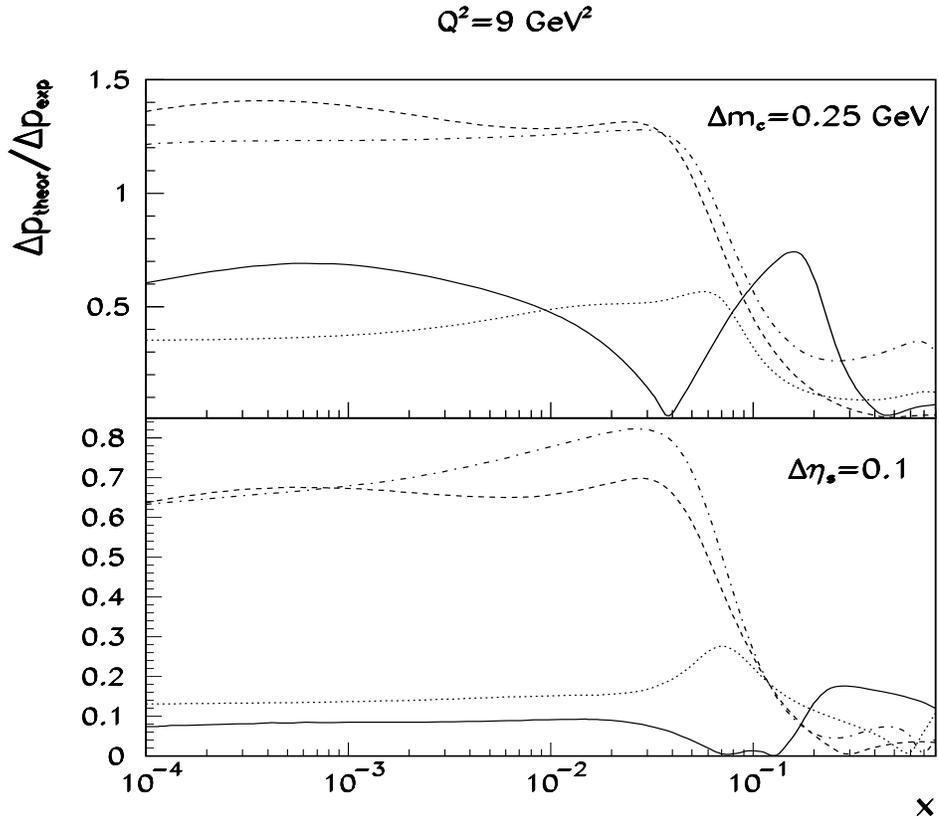,width=14cm,height=12cm}}
\caption{The ratio of magnitude of the uncertainty due to 
errors in the $c$-quark mass (upper panel)
and the strange sea suppression factor (lower panel) to the experimental 
errors in NLO 
PDFs (full lines: gluon distribution; dashes: non-strange sea one;
dots: d-quark one; dashed-dots: u-quark one).}
\label{fig:th}
\end{figure}

This uncertainty is estimated as the variation of PDFs under the change of 
$m_{\rm c}$ from $1.5~{\rm GeV}$ to $1.75~{\rm GeV}$, i.e. within 
the error in its value estimated from other measurements.
The magnitude of corresponding 
uncertainties in PDFs decreases from the LO to the NNLO since the
HO PDFs are smoother at small $x$ and variation of the photon-gluon fusion
cross section gets less sensitive to the shift of the kinematical 
variables. The ratio of its magnitude 
to the PDFs experimental errors in the NLO is given in Fig.\ref{fig:th}.
The ratio is generally less than 1.5 being most important for the 
sea quarks distribution at small $x$. The similar ratio for 
uncertainty due to variation of the $b$-quark mass is negligible for 
all PDFs.

\subsection{Uncertainty due to variation of the strange sea 
parameterization}

The $s$-quark distribution equally contributes to all structure 
functions of the charged leptons DIS and cannot be determined from these 
data due to lack of constraints. Most precise 
information about this distribution comes from the neutrino nucleon DIS data.
Its shape was determined by the NuTeV collaboration to be comparable to 
the average of the $u$- and $d$-quarks distributions
and the total suppression factor of the $s$-quark distribution
$\eta_{\rm s}$ was found to be   
$0.42\pm0.07 ({\rm stat.})\pm0.06 ({\rm syst.})$~\cite{Adams:1999sx}.
We estimate the uncertainty in PDFs due to error in the $s$-quark
distribution as change of the PDFs under 
variation of $\eta_{\rm s}$ from 0.42 to 0.52.
This uncertainty marginally changes from the LO to the NNLO  
and does not exceed 
the PDFs experimental errors (see Fig.\ref{fig:th}).

\subsection{Other theoretical errors}

The list of possible
theoretical errors in PDFs is rather conventional. For example the 
errors due to HT contribution
can be estimated as shift of the fitted PDFs under 
addition to the power-like terms with the $x$-shape motivated by a 
certain model considerations to the fitted function. 
Alternatively, it can be accounted for by fitting possible HT contribution
in the model-independent form; in this case the 
error due to the HT is combined with the total experimental error in PDFs. 
Since the later approach is less subjective we use it in our analysis.
Similarly, we do not treat the error due to the 
uncertainty in value of $\alpha_{\rm s}$ as a separate theoretical error
since we fit $\alpha_{\rm s}$ simultaneously with other parameters. 
In our previous analysis of Ref.~\cite{Alekhin:2001ch}
we considered the error due to uncertainty in the 
heavy-quarks thresholds in the QCD renormalization group
equation for $\alpha_{\rm s}$. Now we omit this error 
because it is generally negligible as compared to the experimental 
errors and other theoretical errors. 
The errors due to uncertainty in the nuclear effects in deuterium, which also 
was considered in Ref.~\cite{Alekhin:2001ch},
will be analyzed in details elsewhere~\cite{AKL}.

\section {High-twist contribution and the value of 
\boldmath{$\alpha_{\rm s}$}}

\begin{figure}
\centerline{\epsfig{file=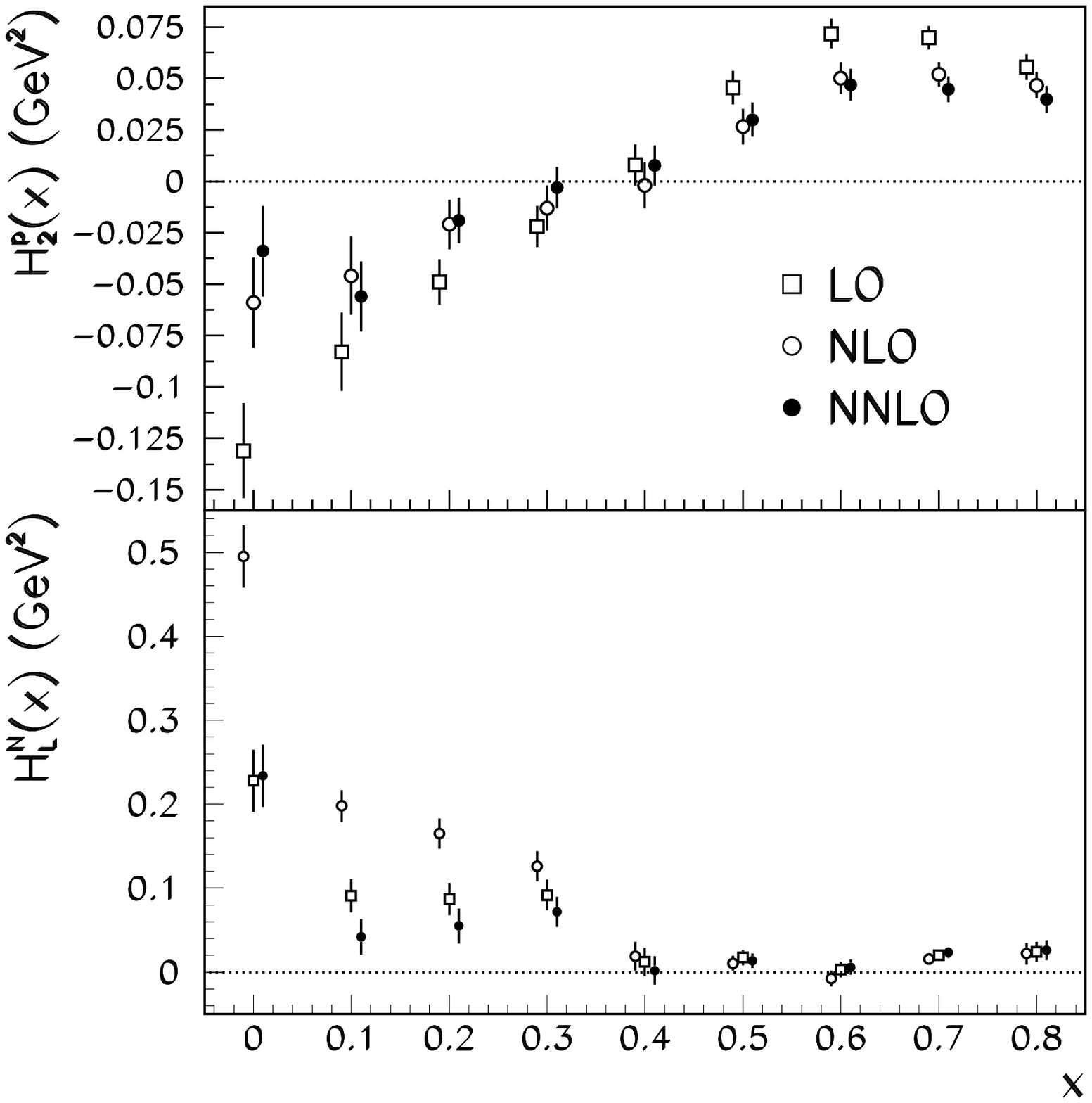,width=14cm,height=12cm}}
\caption{The HT contributions to the 
proton structure functions $F_2$ (upper panel)
and the nucleon structure function 
$F_{\rm L}$ (lower panel) obtained in the different orders of pQCD.
For better view points are shifted to left and right along the $x$-axis.}
\label{fig:ht}
\end{figure}

In the analysis of DIS data with relatively low $Q$ 
one needs to account for the HT terms due to these terms are 
essential part of the operator product expansion \cite{Wilson:zs}.
The HT terms cannot be accurately 
calculated from the existing theory of strong interaction and we have to 
account for them in our analysis phenomenologically.
From comparison with the data we observe non-negligible
contribution from the HT terms 
since account of such terms improves quality of the
fit (e.g. in the NNLO fit the value of $\chi^2$ rises by 330, if we fix the  
HT terms at 0). At the same time the observed HT terms can come 
just from the possible defects of the fitted model, in particular due to
inaccount of the HO QCD corrections. 
Indeed, with the inclusion of the NLO correction 
in the analysis of data on the $\nu N$ DIS the observed 
HT terms decrease and eventually vanish in the NNLO~\cite{Kataev:1999bp}.
We also observe decrease of the HT terms with the pQCD order, but they do 
not vanish completely in our case (see Fig.\ref{fig:ht} and 
Table\ref{tab:ht}).
Extrapolation of the LO-NLO-NNLO results for the HT terms 
allows to expect that they should not vanish in the NNNLO also.
This is in line with the conclusion of Ref.~\cite{Alekhin:1999kt}
that the HO corrections can successfully simulate power behavior 
in narrow region of $x$ close to 0.1 only. This is also in agreement 
with the results of Ref.~\cite{Schaefer:2001uh} (the effects of soft gluon 
resummation considered in Ref.~\cite{Schaefer:2001uh}
are important at $x\gtrsim 0.8$ only
and cannot further decrease the value of HT terms in our case).
We see no contradiction between our results
and the results of Ref.~\cite{Kataev:1999bp} in view of 
large errors in the HT terms extracted from the neutrino data, but
do not support the conclusion about vanishing of the HT terms in the NNLO.
The obtained HT contribution to $F_2$ is maximal at $x\sim 0.6$ 
and at $Q^2=5~{\rm GeV}^2$ is $\sim 10\%$ of the LT term. This is much 
larger than the value of the relative experimental error in the 
$F_2$ measured in this region and hence the account of the 
HT terms is indeed important. 

\begin{table}
\begin{center}
\caption{Fitted coefficients of the 
HT contributions to the structure functions 
$F_{\rm 2,L}$ in the NNLO}
\vspace{0.4cm}
\begin{tabular}{cccc}
$x$ & $H_2^p(x)$ & $H^N_{\rm L}(x)$ \\ \hline 
0. & $-0.034\pm0.022$ & $0.234 \pm 0.037$ \\
0.1 & $-0.056\pm0.017$ & $0.042 \pm 0.021$ \\
0.2 & $-0.019\pm0.011$ & $0.055 \pm 0.021$ \\
0.3 & $-0.003\pm0.010$ & $0.072 \pm 0.018$ \\
0.4 & $0.0077\pm0.0098$ & $0.002 \pm 0.017$ \\
0.5 & $0.0299\pm0.0082$ & $0.0138 \pm 0.0087$ \\
0.6 & $0.0469\pm0.0077$ & $0.0060 \pm 0.0091$ \\
0.7 & $0.0447\pm0.0062$ & $0.0233 \pm 0.0063$ \\
0.8 & $0.0398\pm0.0065$ & $0.026 \pm 0.012$ \\
\end{tabular}
\end{center}
\label{tab:ht}
\end{table}

\begin{table}
\begin{center}
\caption{Values of $\alpha_{\rm s}(M_{\rm Z})$ obtained in different 
orders of pQCD.} 
\vspace{0.4cm}
\begin{tabular}{cc}
LO&   $0.1301\pm0.0026 ({\rm exp})$ \\ 
NLO & $0.1171\pm0.0015 ({\rm exp})$ \\ 
NNLO & $0.1143\pm0.0014 ({\rm exp})$ \\
\end{tabular}
\end{center}
\label{tab:als}
\end{table}

Values of $\alpha_{\rm s}(M_{\rm Z})$ obtained in different orders of pQCD 
are given in Table~\ref{tab:als}. For comparison 
in the NNLO analysis of the existing data for DIS 
of charged leptons off proton the value of
$\alpha_{\rm s}(M_{\rm Z})=0.1166\pm0.0009~({\rm exp})$ was obtained 
\cite{Santiago:2001mh}. This value is comparable with our result, but 
one is to note that, despite the low $Q$ data were used in the fit,  
the analysis of Ref.~\cite{Santiago:2001mh}
was performed with the HT terms fixed at zero. 
Meanwhile the values of $\alpha_{\rm s}$ and $H_2$ at large $x$ 
are strongly anti-correlated. 
From the trial fit with the constraint $H_{2,\rm L}=0$ we obtain 
the value of $\alpha_{\rm s}(M_{\rm Z})=0.1215\pm 0.0003~({\rm exp})$, i.e.
their  value is much larger and their error is much smaller
than in the fit with the HT terms fitted. From this comparison 
we conclude that in fact our result for $\alpha_{\rm s}$ is in disagreement
with the result of Ref.~\cite{Santiago:2001mh}. 
In addition, value of $\alpha_{\rm s}$ obtained in 
the analysis of Ref.~\cite{Santiago:2001mh} rises from the NLO to the NNLO, 
in contrast with our results.

\section{Comparison with other parameterizations}

\begin{figure}
\centerline{\epsfig{file=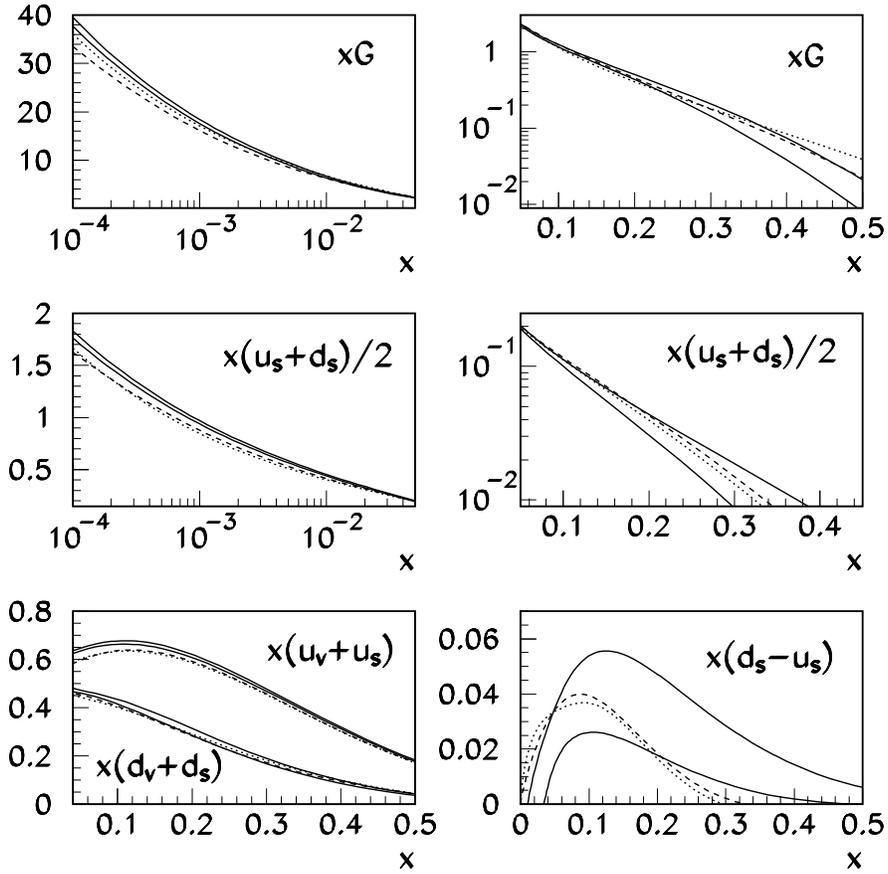,width=14cm,height=14cm}}
\caption{One-standard-deviation
bands for our NLO PDFs evolved in the VFN scheme 
(full lines) compared to the MRST2001 (dashes) and CTEQ6 (dotes) PDFs.} 
\label{fig:pdfcomp}
\end{figure}

Our NLO PDFs evolved in the VFN scheme
are compared to the recent CTEQ and MRST parameterizations in 
Fig.~\ref{fig:pdfcomp}. The comparison is made  
at $Q^2=100~{\rm GeV}^2$, which is the lowest scale 
for the hard processes at the hadron colliders. At this scale we observe 
the largest deviations of our PDFs from the CTEQ gluon distribution
at large $x$, the MRST gluon distribution at small $x$, and the 
quark distribution of both groups at small $x$. 
The observed differences in particular can be  
due to the data on Drell-Yan process and jet production 
were used in the CTEQ and MRST analysis. 
To clarify this point thorough analysis of universality of the PDFs 
in different processes is necessary, but 
any way these differences can have impact on 
the calculated cross sections of hard processes
at the hadron collider, e.g. on the 
K-factor for the Higgs boson production.
The discrepancy with the MRST gluon distribution, which is the lowest of 
the three considered PDFs sets, is especially important
since at small $x$ it gets negative at small $Q$.
At larger $Q$ the difference between the different PDFs sets
is smaller due to focusing properties of the QCD evolution \cite{DeRujula:rf}.

Recently the CTEQ collaboration estimated the errors in their PDFs
and we can compare those estimates with the errors in our PDFs.  
The comparison cannot be completely sensible since 
due to technical difficulties CTEQ did not take into 
account the impact of the normalization errors in data. 
Indeed, the account of normalization errors
essentially increase the point-to-point correlations for 
many subsets and leads to amplification of the numerical instability of the 
calculations. The value of $\Delta R$, which is indicator 
of these correlations (see Sec.\ref{sec:pdfexp}), rises 
essentially if normalization errors are included into the covariance matrix.
In particular, for the BCDMS data $\Delta R$ is 0.68/0.29 with/without 
normalization errors taken into account. 
Meanwhile account of the normalization errors leads to increase of
the PDFs uncertainties up to factor of 5 (see Fig.\ref{fig:pdfsys}).
In the analysis of CTEQ 
the value of $\chi^2/NDP$ for some data subsets deviates
off 1 much more than this is statistically allowed. 
These fluctuations may be caused by several reasons, including 
inaccuracies in the 
model of data, such as inaccount of the HT terms, the target 
mass corrections, the nuclear corrections, and the HO QCD corrections.
To avoid this problem CTEQ scaled their PDFs errors by the 
``tolerance'' factor, which was selected equal to 10 in order to provide 
comparability of all data subsets. Effectively, the application of such 
tolerance factor leads to the crude account of all sources 
of uncertainties, including theoretical ones, since 
the scaled errors cover the total range of possible variation of the 
fitted parameters. In our analysis we need not introduce 
the tolerance factor since deviations of $\chi^2/NDP$ 
for separate data subsets off 1 fit to the possible 
statistical fluctuations (see Table~\ref{tab:chi}).
The errors in CTEQ6 PDFs are compared to ours in Fig.~\ref{fig:err}.
For all PDFs, excluding  the sea quarks distribution at $x\sim 0.3$
the CTEQ PDFs errors are larger than ours.
The difference is especially  pronounced for the gluon and sea
distributions at small $x$ and for the 
$u$- and $d$-quark distributions at large $x$.
From this comparison follows that use of the variety of data for the PDFs
extraction not necessarily leads to the decrease of the PDFs errors 
since improvement in experimental accuracy can be accompanied 
by the growth of theoretical errors (at least one is to take care 
about accurate estimation of the later).
\begin{figure}
\centerline{\epsfig{file=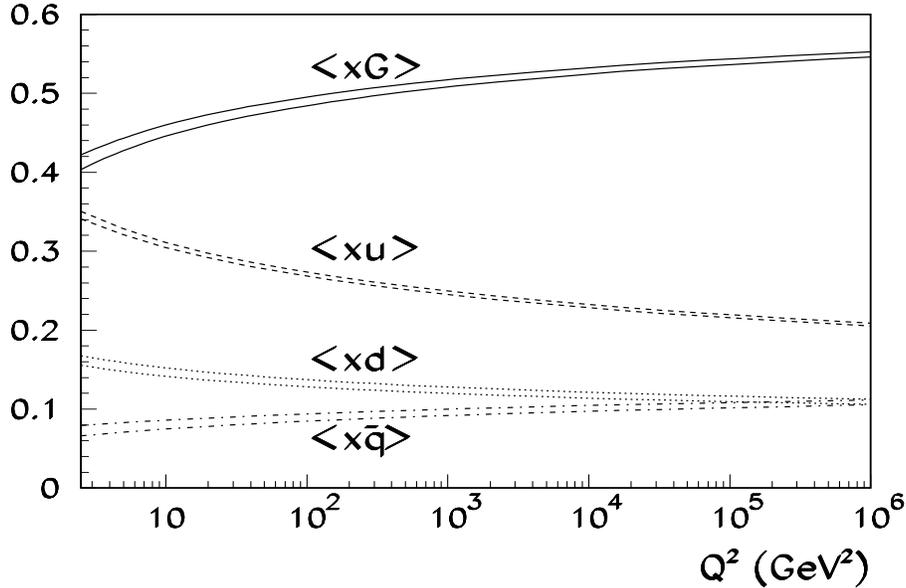,width=14cm,height=9cm}}
\caption{One-standard-deviation bands for the $Q$-dependence of the 
momentum carried by different parton species ($\bar q$ reads
sum of the anti-quark distributions).}
\label{fig:pmom}
\end{figure}

\section {Access to the PDFs} 

\begin{figure}
\centerline{\epsfig{file=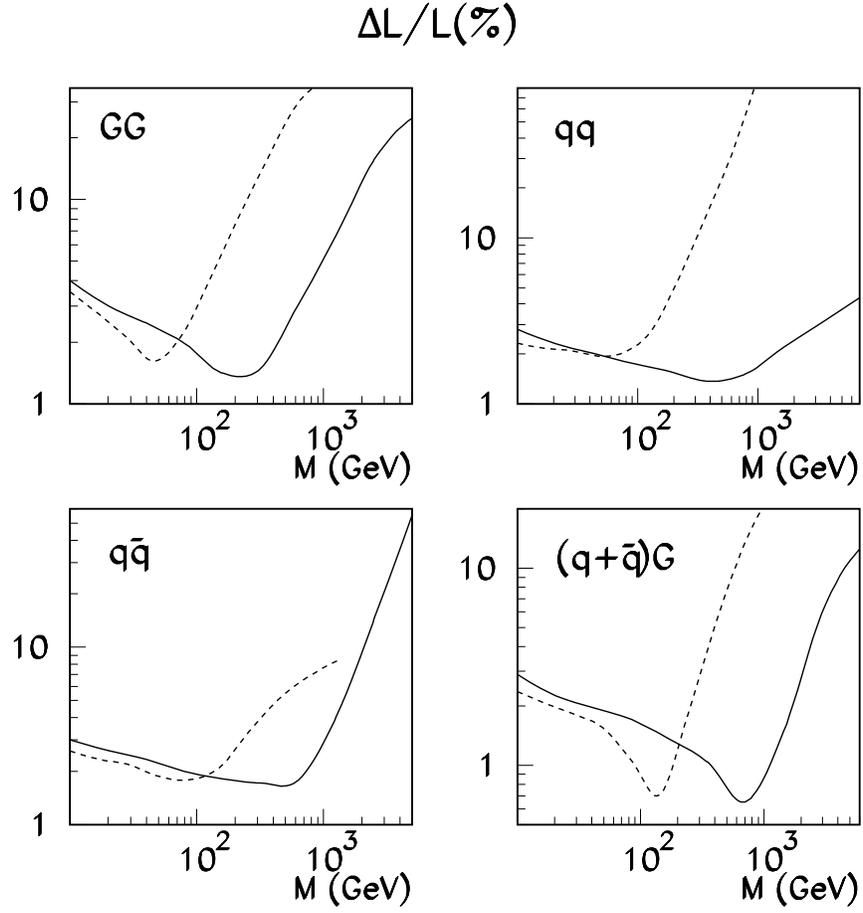,width=14cm,height=14cm}}
\caption{The dependence of 
parton-parton luminosities for the LHC (full curves)
and the Fermilab collider (dashes) on the produced mass $M$
(${\rm qq}:~L_{\rm uu}+L_{\rm dd}+L_{\rm du};~ 
{\rm q\bar q}:~L_{\rm u \bar d}+L_{\rm d \bar u};~ 
{\rm (q+\bar q)G}:~L_{\rm uG}+L_{\rm \bar uG}+L_{\rm dG}+L_{\rm \bar dG}$).}
\label{fig:lum}
\end{figure}

The PDFs obtained in our analysis are available through 
the WWW page\footnote{http://sirius.ihep.su/$\tilde{~}$alekhin/pdfa02/}.
We placed in this page the PDFs generated 
in the LO, NLO, and NNLO; in the VFN and FFN schemes, including special 
sets with shifted values of the $c$-quark mass and the sea suppression 
factor to allow account of the corresponding theoretical uncertainties in PDFs.
The PDFs are supplied by the experimental uncertainties, which 
can be conveniently taken into account in the calculation.
This can be performed both using propagation error formula based on the 
PDFs derivatives and by generating random PDFs. The later way is 
most convenient in the Monte-Carlo calculations and was adopted in 
Les Houches interface for the PDFs with uncertainties \cite{Giele:2002hx}.
The $1\sigma$ bands for the first Mellin moments
of the parton distributions at different $Q^2$ calculated using this code
are given in Fig.\ref{fig:pmom}. The typical uncertainties in 
the momentum carried by different species is generally less
than 0.01, being the largest for the gluon distribution.
The experimental errors in typical parton luminosities $L$
for the antiproton-proton collisions at $\sqrt{s}=2~{\rm TeV}$ 
(Fermilab collider) and the proton-proton 
collisions at $\sqrt{s}=14~{\rm TeV}$ 
(LHC collider) are given in Fig.\ref{fig:lum}.
In the kinematical range relevant for the studies of hard processes 
the relative errors in parton luminosities for the LHC are in the 
range of $1\div 20$\%; for the Fermilab collider they
are comparable with the LHC ones at $M\lesssim~100~{\rm GeV}$ 
and are much larger at $M\gtrsim 100~{\rm GeV}$.
The uncertainty in $W$-boson production 
cross section due to errors in PDFs is
$\sim 2$\% for the both colliders; for the Higgs boson 
production cross section it is $\sim 2$\% for the LHC 
and varies from 2\% to 10\% for the Fermilab collider under variation of 
the Higgs boson mass from $100~{\rm GeV}$ to $300~{\rm GeV}$. 

\section{Conclusion}

The PDFs extracted from the DIS data only are precise enough:  
The estimated uncertainties in the parton luminosities
including all sources of experimental errors in data
are $\lesssim 20\%$ in the kinematical region feasible at  
the LHC collider and $\lesssim 60\%$ at the Fermilab collider.
The theoretical errors in PDFs are under control and generally do not exceed 
experimental ones, therefore the DIS PDFs can be used for 
conclusive searches of deviation from the Standard Model
in other processes and/or checks of the universality of PDFs.

{\bf Acknowledgments}

I am indebted to E.~Boos, M.~Botje, W.~Giele, A.~Kataev, 
T.~Sjostrand, W.K.~Tung, and A.~Vogt for stimulating discussions.
The work was supported by the RFBR grant 00-02-17432.

\end{document}